\documentclass{article}
\usepackage{spconf,amsmath,graphicx}
\usepackage{amsfonts}

\DeclareMathOperator{\MPEG}    {MMPE_G}
\DeclareMathOperator{\MvPEG}    {MMPE_G}
\DeclareMathOperator{\MMPEG}    {MMPE_G}
\DeclareMathOperator{\MMPE}    {MMPE}
\DeclareMathOperator{\PEG}    {PE_G}

\newcommand{\set}[2]{\{ \, #1 \, | \, #2 \, \} } 
\newcommand{\card}[1]{\lvert#1\rvert}
\newcommand{\floor}[1]{\lfloor #1 \rfloor}
\newcommand{\map}[3]{ #1 \colon #2 \longrightarrow #3}

\newcommand{\G}{{G}}
\newcommand{\R}{\mathbb{R}} 
\newcommand{\Id}{\mathbb{I}} 
\newcommand{\V}{\mathcal{V}} 
\newcommand{\E}{\mathcal{E}} 
\newcommand{\Na}{\mathbb{N}}
\newcommand{\A}{\mathbf{A}} 
\newcommand{\Nb}{\mathcal{N}} 
\newcommand{\X}{\mathbf{X}} 

\newcommand{\U}{\mathbf{U}} 
 
\newcommand{\CGE}{\mathcal{G}_{\U^\epsilon}}

\usepackage{enumitem}
\usepackage{etoolbox}
\patchcmd{\thebibliography}
{\settowidth}
{\setlength{\itemsep}{0pt plus 0.1pt}\settowidth}
{}{}
\apptocmd{\thebibliography}
{\small}
{}{}

\title{A noise-robust Multivariate Multiscale Permutation Entropy for two-phase flow characterisation}
\name{John~Stewart~Fabila-Carrasco$^{\star 1}$ \qquad Chao Tan$^{\dagger}$ \qquad Javier Escudero$^{\star 2}$\thanks{J.S. Fabila-Carrasco and J. Escudero were supported by the Leverhulme Trust via a Research Project Grant (RPG-2020-158).}}

\address{$^{\star}$ University of Edinburgh, UK. \quad \emph{$^{1}$ john.fabila@ed.ac.uk} \quad \emph{$^{2}$ javier.escudero@ed.ac.uk}\\
	$^{\dagger}$ Tianjin University, China. \emph{tanchao@tju.edu.cn}\\
}

\begin{document}
	\setlength{\abovedisplayskip}{0pt}
	\setlength{\belowdisplayskip}{0pt}
	%
	\maketitle
	\begin{abstract}
Using a graph-based approach, we propose a multiscale permutation entropy to explore the complexity of multivariate time series over multiple time scales. This multivariate multiscale permutation entropy ($\MPEG$) incorporates the interaction between channels by constructing an underlying graph for each coarse-grained time series and then applying the recent permutation entropy for graph signals. Given the challenge posed by noise in real-world data analysis, we investigate the robustness to noise of $\MPEG$ using synthetic time series and demonstrating better performance than similar multivariate entropy metrics. 

We also apply $\MPEG$ to study two-phase flow data, an important industrial process characterised by complex, dynamic behaviour. To this end, we process multivariate Electrical Resistance Tomography (ERT) data and extract multivariate multiscale permutation entropy values. $\MPEG$ characterises the flow behaviour transition of two-phase flow by incorporating information from different scales. The experimental results show that $\MPEG$ is sensitive to the dynamic of flow patterns, allowing us to distinguish between different flow patterns. We show that our method is noise-robust, which is suitable for analysis of the complexity of multivariate time series and characterising two-phase flow recordings.
	\end{abstract}
	\section{Introduction}
	\label{sec:intro}
Two-phase flow is important in many industries, including chemical processes, petroleum exploitation, nuclear engineering, and transportation \cite{levy1999two}. Theoretical and experimental studies have been sought to characterise two-phase flow, including mathematical approaches~\cite{drew1983mathematical, stewart1984two}, fluid dynamics analysis~\cite{ishii2010thermo}, using high-speed camera \cite{hongwei2010multi}, and others. These methods have not fully resolved the complexity and dynamic behaviours of the flow patterns, especially in the interaction between different channels in the multivariate signals.

The analysis of a complex system like two-phase flow can benefit from nonlinear analysis metrics. Univariate permutation entropy~\cite{Bandt2002} has been used as a nonlinear measure of complexity; it is a computationally fast algorithm, its performance under noise conditions has been investigated~\cite{Veisi2007}, has been used for discriminating two-phase flow dynamics~\cite{Fan2013} and characterise autoregressive processes~\cite{davalos19}. Most of the physical systems' signals are multivariate. Therefore, several univariate entropy metrics have been generalised to a multivariate setting, including multivariate sample entropy~\cite{Ahmed2011}, multivariate dispersion entropy ~\cite{Azami2019}, and multivariate permutation entropy~\cite{Morabito2012a}, among others. Some of them have been recently used to analyse phase flow and are applied to characterise the behaviour of the two-phase flow system~\cite{Tan2013, Gao2015,gao2016}. A prior implementation of multivariate permutation entropy exists, but it has a major limitation (i.e., it does not consider cross-channel information).

The effect of noise on the multivariate permutation entropy can lead to inaccurate values. Some improvements have been proposed to deal with this problem, such as a multivariate weighted permutation entropy~\cite{Yin2017}. However, these multivariate methods analyse each time series separately; hence, the cross-channel information will be lost and increase the number of parameters used for the entropy computation. To this end, we present a multiscale multivariate permutation entropy based on constructing a Cartesian graph product.
	
\emph{Contributions:} This paper introduces a multiscale algorithm to analyse multivariate time series based on the permutation entropy for graph signals. We apply the algorithm to a set of synthetic data and two-phase flow data. We show that it improves the performance of univariate permutation entropy and classical multivariate permutation entropy because $\MPEG$ consider the interaction between the different data channels, it is robust to noise, and it is useful to detect the complexity at different scales of the phase flow patterns.

\emph{Structure of the manuscript:} The outline of the manuscript is as follows: Sec.~\ref{back} introduces the permutation entropy for graph signals and Cartesian graph product. Sec.~\ref{multiscale} presents the multiscale multivariate permutation entropy $\MPEG$. In Sec.~\ref{noise}, we use synthetic signals to show $\MPEG$ is robust to noise and it is applied to analyse phase flow data (Sec.~\ref{flow}). The conclusions and future work are presented in Sec.~\ref{res}.	

	\section{Graphs and permutation entropy }\label{back}
	
This section presents the Cartesian graph product and the recently permutation entropy for analyse graph signals.

\textbf{Cartesian graph product.} 
The \emph{Cartesian product} of two graphs $\G= (\V,\E)$ and $\G'=(\V',\E')$, denoted $\G\square \G'$, is the graph defined by the vertex set: $\V(\G\square \G')=\V\times \V'=\set{(v,v')}{v\in\V \text{ and } v'\in\V'}\:$. Two vertices $(v,v')$ and $(u,u')$ are adjacent in $\G\square \G'$ if and only if either $v=u$ and $v'$ is adjacent to $u'$ in $\G'$, or $v'=u'$ and $v$ is adjacent to $u$ in $\G$. The graph product is a useful structure to model multidomain signals~\cite{kadambari2020learning} and it is the perturbation of a periodic graph~\cite{phd2020}.

\textbf{Permutation entropy for graph signals ($\PEG$).} 
Let $G = (\V,\E)$ be a graph, $\A$ its adjacency matrix and $\textbf{X}=\left\{x_i\right\}_{i=1}^{n}$ be a signal on the graph, $\PEG$ is defined in~\cite{Fabila2021,Fabila2022} as follows: 
\begin{enumerate}[wide, labelwidth=!, labelindent=0pt, noitemsep]
	\item  For $2\leq m\in\Na$ the \emph{embedding dimension}, $L\in\Na$ the \emph{delay time} and for all $i=1,2,\dots,n$, we define 
	$y_{i}^{kL}= \frac 1 {\card{\Nb_{kL}(i)}} \sum_{j \in \Nb_{kL}(i)} x_j=\frac{1}{\card{\Nb_{kL}(i)}}(\A^{kL}\X)_i\;$, where
	$\Nb_k(i)=\set{j\in \V}{\scalebox{.9}[1.0]{it exists a walk on $k$ edges joining $i$ and $j$}}\;$. Then, we construct the embedding vector $\textbf{y}_i^{m,L}\in\R^m$ given by \[\textbf{y}_i^{m,L}=\left( y_{i}^{kL}\right)_{k=0}^{m-1}=\left(y_i^0,y_{i}^{L},\dots y_{i}^{(m-1)L}\right)\;.\]
	\item The embedding vector $\textbf{y}_i^{m,L}$ is arranged in increasing order and is assigned to one of $k= m!$ permutation (or patterns) $\pi_1,\pi_2,\dots,\pi_k$.
	\item For the distinct permutation, the relative frequency is denoted by $p(\pi_1),p(\pi_2),\dots,p(\pi_k)$. The permutation entropy $\PEG$ for the graph signal $\textbf{X}$ is computed as the normalised Shannon entropy
	\begin{equation*}
	\PEG=-\dfrac{1}{\ln(m!)}\sum_{i=1}^{m!} p(\pi_i) \ln p(\pi_i)\;.
	\end{equation*}
\end{enumerate}
\section{Multiscale Multivariate Permutation Entropy ($\MMPEG$)}\label{multiscale}
Time series from phase flow data contain multiple temporal scale structures. Consider only a single scale have limited capability and they can only assess the system's irregularity at a single temporal scale. Multiple scales need to be analysed to understand the dynamics in the signals and to describe the properties of its model.

Here, we propose a nonlinear multivariate multiscale methodology based on graph signals to analyse such data. 
Let $\X=\{x_{t,s}\}_{t=1,2,\dots,n}^{s=1,2,\dots,p}$ be a multivariate time series with $p-$channels of length $n$ and with $I_p$ the graph of interactions between channels. The $\MPEG$ algorithm relies on a three-step procedure: 
\begin{enumerate}[wide, labelwidth=!, labelindent=0pt, noitemsep]
	\item \textbf{Coarse-grained procedure.} From the original univariate signal $\X$, we derive multiple successive coarse-grained versions by averaging the time data points within non-overlapping time segments of increasing length, $\epsilon$ , referred to as the scale factor. Each element of the coarse-grained time series, $\U^\epsilon=\{u_{i,j}^\epsilon\}_{i=1,2,\dots,\floor{n/\epsilon}}^{j=1,2,\dots,p}$ , is calculated as:
	\begin{equation*}
		u_{i,j}^\epsilon=\dfrac{1}{\epsilon} \sum_{t=(i-1)\epsilon+1}^{i \epsilon} x_{t,j}\;, \quad \text{for } 1\leq j \leq p\text{ and } 1\leq i\leq n/\epsilon\; .
	\end{equation*}
	The length of each coarse-grained time series is $\epsilon$ times shorter than the original one. For $\epsilon=1$, we get the original series, i.e, $\U^1=\X$.
	
	We use this coarse-grained procedure for simplicity, however other approaches to construct the coarse-graining in multiscale entropy exist~\cite{Azami2018,Valencia2009}.
	\item \textbf{Graph construction associated to a multivariate signal.} For each coarse-grained multivariate $\U^\epsilon$, we construct the graph $\CGE$ given by:
	\[\CGE:=\overrightarrow{P}_{\floor{n/\epsilon}}\square I_p\;.\]
	
	Usually, we will consider two basic graph structures, $I_p=K_p$ the complete graph with $p$ vertices and $I'_p=\emptyset_p$ the empty graph on $p$ vertices i.e., without channels-interaction (see Fig.~\ref{figgraphproduct} for an example with $p=3$).
	\begin{figure}[htb]
	\centerline{\includegraphics[width=85mm]{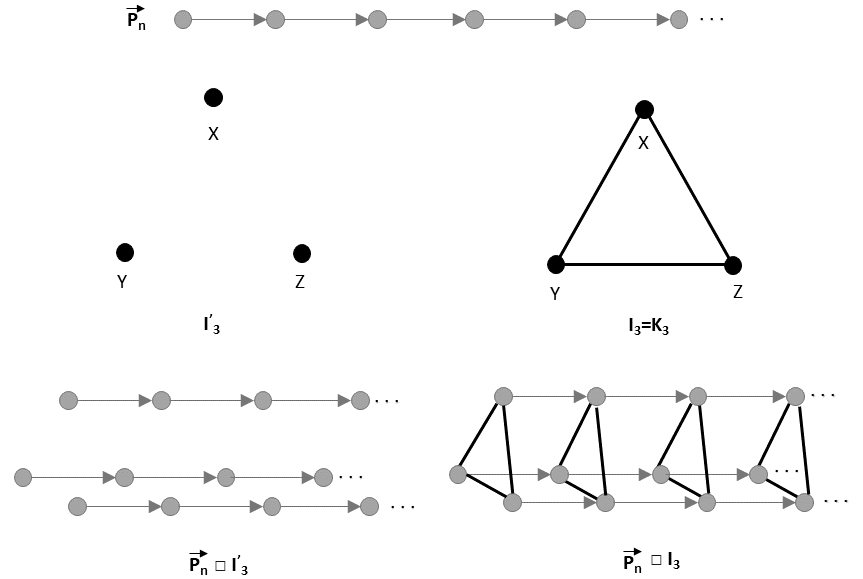}}
	\caption{Graph structures considering as interaction graphs.}
	\label{figgraphproduct}
\end{figure} 	

	The directed graph $\overrightarrow{P}_n$ has $n$ vertices with adjacency matrix ${\mathbf  A}_{\overrightarrow{P}_n}$ (size $n\times n$), and the graph $I_p$ has $p$ vertices with adjacency matrix ${\mathbf  A}_{I_p}$ (size $p\times p$), then the adjacency matrix ${\mathbf  A}_{{\overrightarrow{P}_n}\square I_p}$  (size $np\times np$) of the Cartesian product of both graphs ${\overrightarrow{P}_n}\square I_p$ is given by
	\begin{equation*}
		{\mathbf  A}_{{\overrightarrow{P}_n}\square I_p}=\mathbf {A}_{{\overrightarrow{P}_n}}\otimes \Id _{p}+\Id _{n}\otimes \mathbf {A} _{I_p}\;;
	\end{equation*}
	where $\otimes$  denotes the Kronecker product of matrices and $\Id_{n}$ denotes the $n\times n$ identity matrix. 
	\item \textbf{$\mathbf{PE}$ for graph signals.} We consider $\U^\epsilon$ as a signal defined on the graph $\CGE$, i.e.,
	$\map{\U^\epsilon}{\V(\CGE)}{\R}\;.$ The \emph{multivariate permutation entropy ($\MvPEG$)} is defined as the permutation entropy $\PEG$ for the graph signal $\U^\epsilon$ defined on the graph $\CGE$, i.e., 
	\begin{align*}
		\MvPEG=\PEG(\U^\epsilon)\;.
	\end{align*} 
\end{enumerate} 

To be noted, if we consider $I_p=\emptyset$, our method leads to the multiscale multivariate permutation entropy previously presented in~\cite{Morabito2012a}. However, $\MPEG$ uses information on between channel interaction (represented in the graph $I_p$) and leads to a more robustness method in the presence of noise, as we will see in the next section. 
	\section{Robustness to noises}\label{noise}
\begin{figure}[htb]
		\centerline{\includegraphics[width=8.5cm]{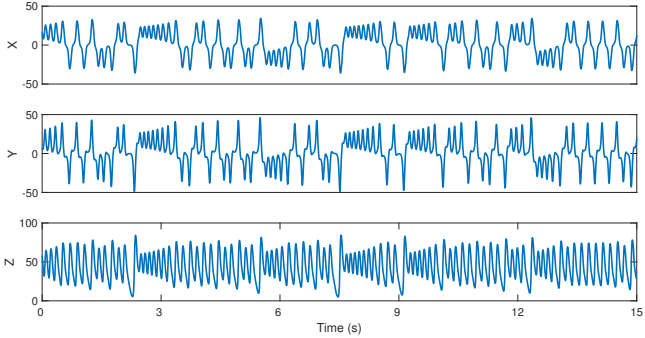}}
	\caption{Time series or sequences of Lorenz system.}
	\label{figLorenz1}
	\end{figure}

	To demonstrate the effectiveness of the method, we will analyse the effect of additive Gaussian noise on the performance of $\MMPEG$ on the Lorenz system, this system has important applications in mechanics, biology, and circuit theory~\cite{Lorenz}. The Lorenz system is is given by the system of ordinary differential equations:
	\begin{align*}{x'}&=\sigma (y-x),\\{y'}&=x(\rho -z)-y,\\{z'}&=xy-\beta z.\end{align*}
	A simulation for the values $\rho=50, \sigma=19$ and $\beta=6$, with initial state $x=-1, y=0$ and $z=1$ is depicted in Fig.~\ref{figLorenz1}. 

\begin{figure}[htb]
	\centerline{\includegraphics[width=81mm]{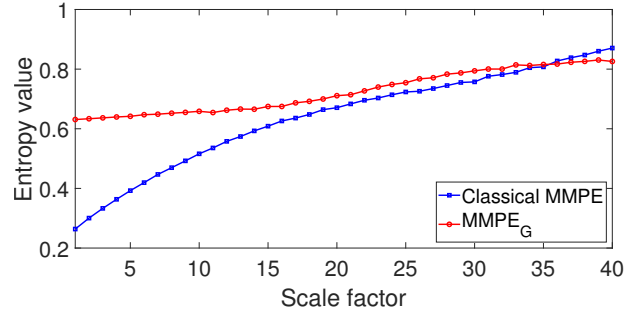}}
	\caption{Entropy values as a function of the temporal scale $\epsilon$.}
	\label{figLorenzentropy}
\end{figure}

We calculate the multivariate permutation entropy, considering the two previous graph structures, $I_3$ and $I'_3$, and we plot the entropy values as a function of the scale factor $\epsilon$ (see Fig.~\ref{figLorenzentropy}). In both graph structures, the entropy values increase with the scale factor. However, our $\MPEG$ shows more stability to the change of scale factor, and the complexity of the multivariate system is more constant, while $\MMPE$ ~\cite{Morabito2012a,Yin2017} shows more variability in its values, changing from no complexity at lower scales to almost random behaviour at higher scales for the same Lorenz system. This robustness would be important for noise.

\begin{figure}[htbp]
	\centerline{\includegraphics[width=85mm]{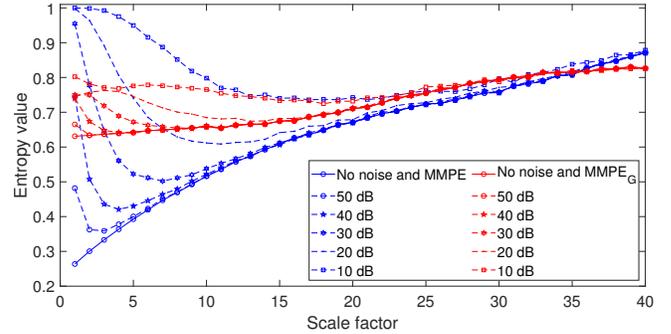}}
	\caption{Entropy values when WGN is added.}
	\label{figrobustnoise}
\end{figure}

To demonstrate the robustness of the method, we will add white Gaussian noise to the multivariate signals defined by the Lorenz system. Fig.~\ref{figrobustnoise} shows the effects of WGN on multivariate multiscale $\MMPE$ and $\MPEG$  for the Lorenz system. The signal-to-noise ratio of WGN is set for all the channels as 10dB, 20dB, 30dB, 40dB and 50dB, respectively. 
The noise has low influence with both methods for scales higher than ten. The influence of the noise is very important for $\MMPE$ in lower scales, while the impact is smaller for our algorithm, showing the robustness of $\MMPEG$. Similar results are obtained when the noise is added to one or two channels.

The analysis of dynamics with $\MPEG$ includes temporal information and channel interaction. Experiments where a noise channel is added to the Lorenz system confirmed that $\MPEG$ has a good performance. The entropy profile did not change noticeably with the addition of the noise channel (results not shown due to space constraints). Preserving all channels (without deleting noise) is important to compare entropy values for different systems but with the same number of channels. 
\section{Flow analysis with $\MMPEG$}\label{flow}
The two-phase flow experiment was carried out at Tianjin University~\cite{Tan2013}, using Electrical Resistance Tomography (ERT). Based on the principle that the conductivity of medium differs, ERT collects boundary voltages between electrodes placed around the pipe by applying electric currents to obtain conductivity distribution of two-phase flow inside the pipe. A constant electrical current of $50$ kHz is adopted as the exciting signal, and the data acquisition rate is 120 frames/s. A $16-$electrode ERT obtains $16\times 13=208$ voltage data. Given the redundancy in the measurements and similar to~\cite{tan2018gas}, we extract feature vectors $V_{Ri}$ from each electrode to reduce the dimension as follows: $V_{Ri}=\frac{1}{13}\sum_{j=1}^{13}(V_{ij}-V_{i{j_0}})/V_{i{j_0}}$ where $V_{ij}$ is the measure voltage value, $V_{i{j_0}}$ is the $V_{ij}$ when the pipe is full of water and $1\leq R \leq 16$. Finally, to reduce the computational time, the 16 features vectors $V_{Ri}$ are compressed into 4 time series, by average  four $V_{Ri}$ electrode belonging to the same set group~\cite{Tan2013}. Then, we applied multivariate multiscale permutation entropy to characterise the two-phase flow. 
\begin{figure}[htb]
	\centerline{\includegraphics[width=85mm]{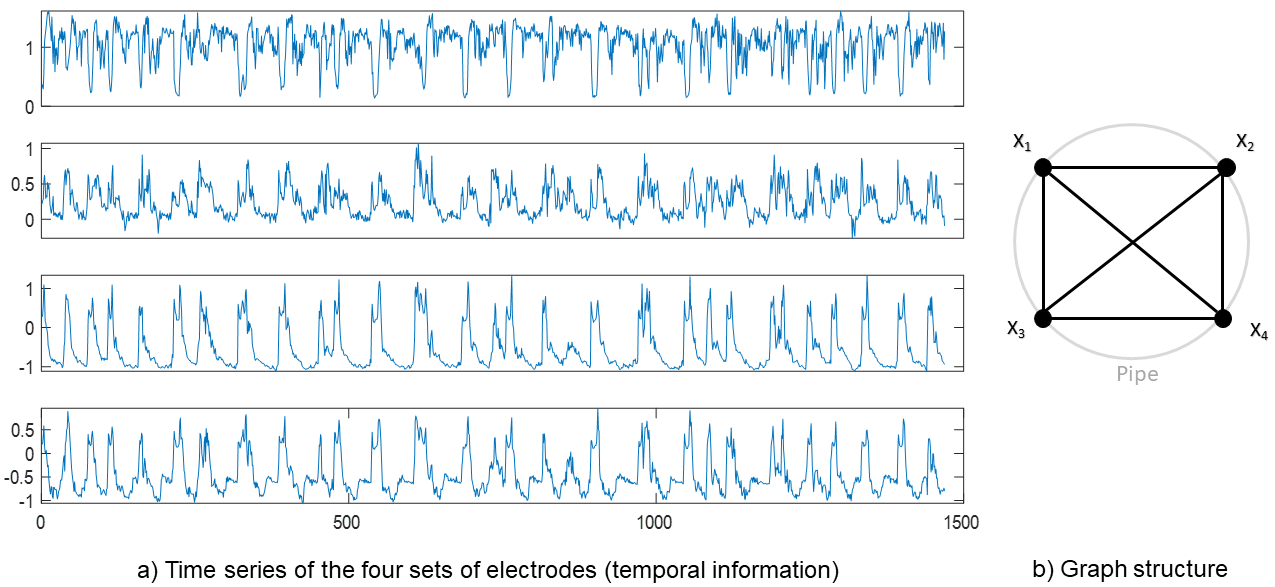}}
	\caption{a) Signals of $4$ set of electrodes under annular flow and b) is the graph modelling cross-channel interaction.}
	\label{fig4timeseries}
\end{figure}

Two-phase flow series require preprocessing to eliminate the noise, but our method is noise-robust as shown in Sec.~\ref{noise} and has better results than classical $\MMPE$. Hence we do not need filtering to obtain a good characterisation of the dynamic in phase flow. The noise causes only small variations of entropy values in the lower scales; hence, we will work with the original data without additional preprocessing.

\begin{figure}[htb]
	\includegraphics[width=85mm]{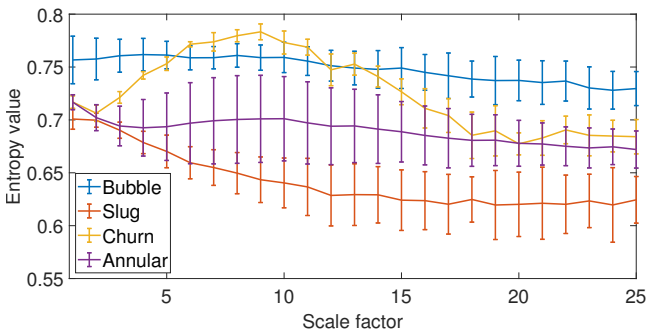}
	\caption{Mean and standard deviation from $\MPEG$ values of signals for different flow patterns.}
	\label{flow_graph}
\end{figure}

For two-phase flow, gas and water were mixed. Water velocity ranged from $0.4$ m/s to $2.9$ m/s and gas velocities from $0.06$ m/s to $5.64$ m/s. The flow pattern observed with this experiments and analysed are patterns known as Bubble flow, Slug flow, Churn flow, and Annular flow. For the analysis, $89$ experiments carried out under different conditions of flow rate of gas and water and the typical length of each recording is $\sim$1400 time samples.  We perform the $\MPEG$ on the signals of the four flow patterns, and plot the corresponding $\MPEG$ versus scale factor in Fig.~\ref{flow_graph}. The coarse-grain process reduces the time series length; consequently, the results show more variability in high-scale factors.

\textbf{Entropy of flow patterns.}
The bubble flow shows the highest $\MPEG$ value in almost all scales compared with the other three patterns. The presence of small bubbles~\cite{dong2017gas} on the regime produces complex time series on all four sets of nodes. Because of gravity, the bubbles affect more the top electrodes than the bottom. Then, the high complexity of the time series and different response of the sensors leads to the highest $\MPEG$ values.

The signals acquired in slug flow show a periodic fluctuation induced by the repeated occurrence of big gas bubbles. Moreover, the length of the gas bubble could be visually identified in the signal because the four electrode sets are all affected by the big gas bubbles flowing over the measured cross-section and show high voltages resulting from the bubble. Hence, the periodicity on temporal dimension and similar effects in all the electrodes produce the lowest $\MPEG$ values for all scales.

The symmetrical distribution of the liquid film around the pipe perimeter in the annular flow leads to the electrodes in the annular flow showing similar fluctuations~\cite{tan2015kalman}, but with different amplitudes. The liquid film at the bottom of the pipe is thicker than the top due to the gas velocity; see Fig.~\ref{fig4timeseries}. The complexity depends more on the temporal dimension than the structural dimension. Hence, the values of $\MPEG$ overlap in low scales for slug and annular flow, making them indistinguishably but resulting in less complexity than bubble flow. For higher scale, $\MPEG$ is able to distinguish between the slug, annular, and bubble flow.    

Churn flow is the flow with more dynamic changes along the scale values because it is a highly unstable flow~\cite{gao2016}. In lower scales, churn flow is similar to the slug and annular flow. The presence of bubbles in the churn flow and the interaction and coalesce with each other produces the highest entropy values than slug and annular flow for scales between 3-14 and more complexity than bubble for scales between 7-11 because of the presence of no complete slugs and small waves. Periodic waves are relevant for higher scales, decreasing the values of $\MPEG$ and making them similar to the annular flow.   

\section{Conclusions and future work}\label{res}
This paper proposes a multiscale nonlinear methodology to analyse multivariate time series using the concept of permutation entropy. Contrary to the previous state of the art, our method allows for the consideration of cross-channel interactions, thanks to the exploitation of graph products and our recent formulation of permutation entropy for graph signals. $\MPEG$ is robust with respect to additive noise, making it suitable for analysis of the complexity of multivariate time series and characterisation of two-phase flow recordings. 

Some future lines of research are: additional statistical analysis of $\MPEG$ values (for example, analysis of the slopes in the scale factor), to explore other underlying graph construction (tensor product, strong product), and to consider other types of data as well, including biomedical signals.

	\vfill\pagebreak
	\bibliographystyle{IEEEbib}
	\bibliography{refs}
	
\end{document}